\documentclass[sigconf]{acmart}

\usepackage{graphicx}
\usepackage{amsmath}
\usepackage{booktabs}
\usepackage{algorithm}
\usepackage{algorithmic}
\usepackage{amsfonts}
\usepackage{multirow}
\usepackage{makecell}
\usepackage{subfigure}
\usepackage{color}
\usepackage{bm}
\usepackage{epstopdf}
\usepackage{url}
\usepackage[cal=cm]{mathalfa}
\usepackage{balance}
\usepackage{threeparttable}
\usepackage{lipsum}
\usepackage{enumitem}
\usepackage{makecell}

\AtBeginDocument{%
  \providecommand\BibTeX{{%
    \normalfont B\kern-0.5em{\scshape i\kern-0.25em b}\kern-0.8em\TeX}}}

\begin{document}
\title{HoME: Hierarchy of Multi-Gate Experts for\\Multi-Task Learning at Kuaishou}
\renewcommand{\shorttitle}{HoME}

\author{Xu Wang}
\affiliation{
  \institution{Kuaishou Technology}
  \country{wangxu28@kuaishou.com}
 }

 \author{Jiangxia Cao}
\affiliation{
  \institution{Kuaishou Technology}
  \country{caojiangxia@kuaishou.com}
 }

 \author{Zhiyi Fu}
\affiliation{
  \institution{Kuaishou Technology}
  \country{fuzhiyi@kuaishou.com}
 }

 \author{Kun Gai}
\affiliation{
  \institution{Unaffiliated}
  \country{gai.kun@qq.com}
 }

 \author{Guorui Zhou}
\affiliation{
  \institution{Kuaishou Technology}
  \country{zhouguorui@kuaishou.com}
 }

\begin{abstract}
In this paper, we present the practical problems and the lessons learned at short-video services from Kuaishou.
In industry, a widely-used multi-task framework is the Mixture-of-Experts (MoE) paradigm, which always introduces some shared and specific experts for each task and then uses gate networks to measure related experts' contributions.
Although the MoE achieves remarkable improvements, we still observe three anomalies that seriously affect model performances in our iteration:
(1) \textbf{Expert Collapse}: We found that experts' output distributions are significantly different, and some experts have over 90\% zero activations with ReLU, making it hard for gate networks to assign fair weights to balance experts. 
(2) \textbf{Expert Degradation}: Ideally, the shared-expert aims to provide predictive information for all tasks simultaneously. Nevertheless, we find that some shared-experts are occupied by only one task, which indicates that shared-experts lost their ability but degenerated into some specific-experts.
(3) \textbf{Expert Underfitting}: In our services, we have dozens of behavior tasks that need to be predicted, but we find that some data-sparse prediction tasks tend to ignore their specific-experts and assign large weights to shared-experts. The reason might be that the shared-experts can perceive more gradient updates and knowledge from dense tasks, while specific-experts easily fall into underfitting due to their sparse behaviors.

Motivated by those observations, we propose HoME to achieve a simple, efficient and balanced MoE system for multi-task learning.
Specifically, we conduct three insightful modifications:
(1) \textbf{Expert normalization\&Swish mechanism} to align expert output distributions and avoid expert collapse.
(2) \textbf{Hierarchy mask mechanism} to enhance sharing efficiency between tasks to reduce occupancy issues and away from expert degradation.
(3) \textbf{Feature-gate\&Self-gate mechanisms} to ensure each expert could obtain appropriate gradient to maximize its effectiveness.
To our knowledge, this paper is the first work to focus on improving multi-task MoE system stability, and we conduct extensive offline\&online (average improves \textbf{0.52\% GAUC} offline \& \textbf{0.954\% play-time per user} online) experiments and ablation analyses to demonstrate our HoME effectiveness.
HoME has been deployed on Kuaishou's short-video services, serving 400 million users daily.
\end{abstract}



\maketitle

\section{Introduction}

\begin{figure}[t!]
\begin{center}
\includegraphics[width=9cm,height=4.3cm]{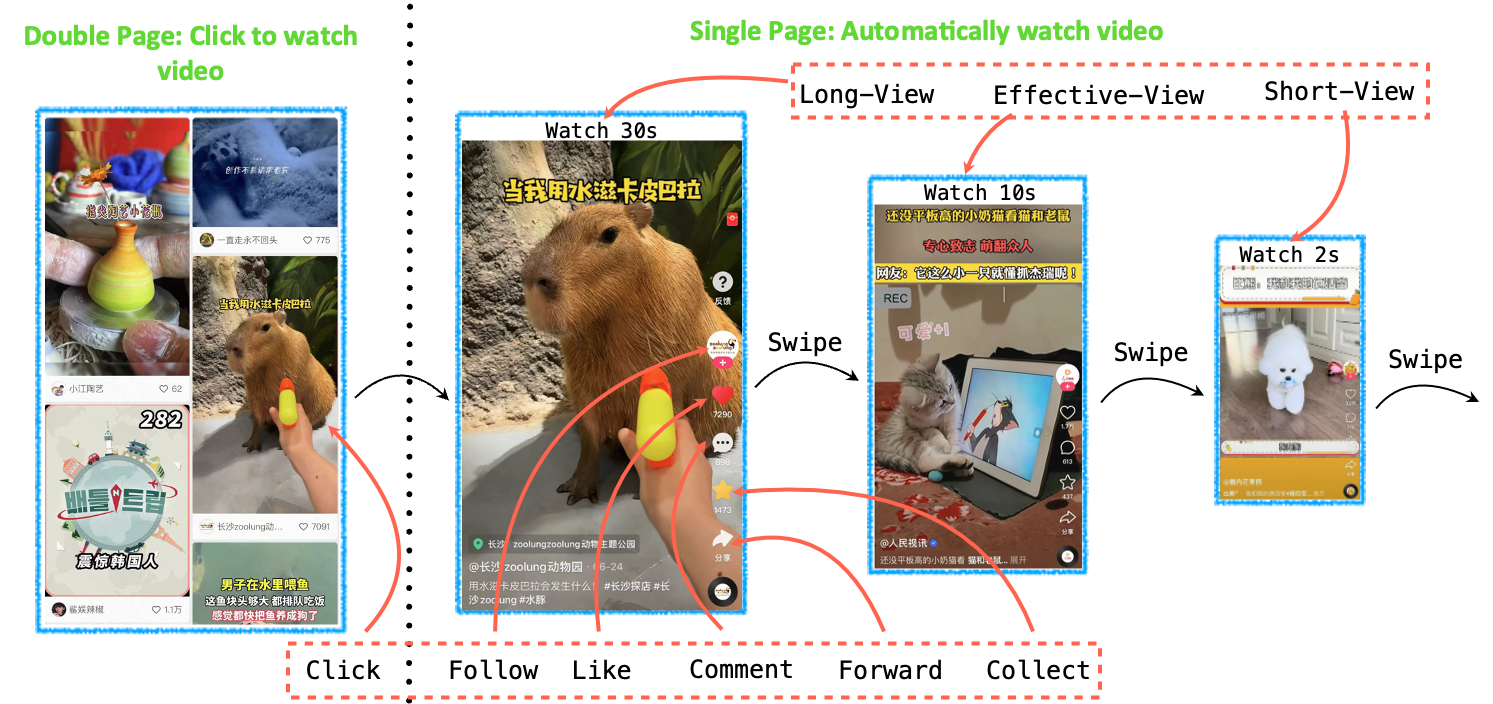}
\caption{Typical multi-task behaviors at Kuaishou.}
\label{kuaishou}
\end{center}
\end{figure}

\begin{figure*}[t!]
\begin{center}
\includegraphics[width=18cm,height=4cm]{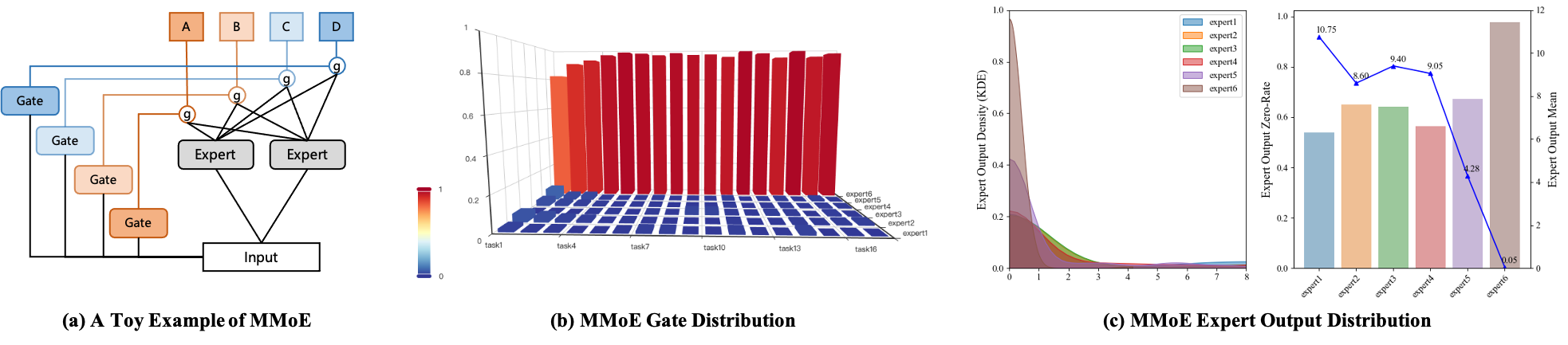}
\caption{Illustration of a naive MMoE and the expert collapse issue occurring in practice.
As shown in (b), expert6 always assigned the biggest gate value, over 0.98 in most cases, by all tasks. 
We also noticed that expert6 outputs much more smaller and sparser activation values than other experts, as shown in (c). Those phenomena indicate that in the real data-streaming scenario, MMoE is unstable and easy to collapse, which obstacles fair comparisons among experts and impacts model performance.
}
\label{mmoe}
\end{center}
\end{figure*}

Short-video applications like Tiktok and Kuaishou have grown rapidly in recent years; unlike other platforms, users always have clear intentions, i.e., search keywords on Google and buy clothes/food at Amazon, while Kuaishou almost plays an entertainment role without any users' concept inputs.
As shown in Figure~\ref{kuaishou}, when using Kuaishou, users usually watch multiple automatically played short-videos by simply swiping up and down on the screen, and sometimes leave some interactions, e.g., Long-view, Comment, etc.
The proportion of implicit feedback~\cite{dfn, gong2022positive} is much greater than other scenarios. Therefore, the only reason why Kuaishou could grow to a large application with 400 million users worldwide, is that our system can provide personalized and interesting short videos, giving users a satisfactory experience.
To this end, utilizing the rare but multifarious behavior cues left by users to capture their interests accurately is the fundamental task.
Generally, the common wisdom always forms such learning process as a \textit{multi-task learning} paradigm~\cite{MTLSurvey, stem, dtrn}, to build a model that could output multiple estimated probabilities of different user interactions simultaneously and supervise this model by real user behavior logs.

As a typical multi-task solution, the idea of MoE is widely used in industry to implement parameter soft-sharing. The most famous method is Multi-gate Mixture-of-Experts (MMoE~\cite{mmoe}), which consists of two major components (as shown in Figure~\ref{mmoe}(a)):
\textit{Expert Networks} -- a group of expert networks (e.g., MLP with ReLU) for modeling the input features and implicit high-level feature crossing as multiple representations, and \textit{Gate Networks} -- task-specific gate networks (e.g., MLP with Softmax) for estimating different experts' importance to fuse their outputs for corresponding tasks.
Recently, several works have extended the expert networks to enhance MMoE system ability by introducing the task-specific experts (e.g., CGC~\cite{ple}) or stacking more experts layers (e.g., PLE~\cite{ple}, AdaTT~\cite{adatt}). 
At Kuaishou, our former online multi-task module is equipped by MMoE~\cite{mmoe}, which remarkably improves our A/B test metrics compared to the baseline.
However, after launching the MMoE, we have tried several different changes to the multi-task modeling module in past years. But all ended in failure, including upgrading to two or more expert layers, extending more shared-experts, introducing extra specific-experts, and so on.
Consequently, we started in-depth analyses to identify the potential reasons that might prevent our iterating.
Unsurprisingly, we discovered three anomalies that seriously affect multi-task performances.

    \textbf{Expert Collapse}: 
    We first checked the gate output situation of MMoE and showed major tasks' gate weight assigned to 6 shared-experts in Figure~\ref{mmoe}(b).
    It is noticeable that all gates assigned larger weights to the shared-expert 6 and almost ignored other shared-experts.
    Thus, we next checked the output value distribution of the shared experts and observed their significant differences. 
    As shown in Figure~\ref{mmoe}(c), the mean and variance of experts 1$\sim$5 are at a similar level, but expert 6 is 100x smaller in terms of the mean value. 
    Such inconsistent output distributions result in the gate network making it difficult to assign fair weights to balance different experts, which further leads the experts at different numerical levels to be mutually exclusive.
    Moreover, we also found that expert output has too many 0 activations (i.e., over 90\% of output), causing its average derivatives to be small and parameters insufficiently trained.

\begin{figure}[t]
\begin{center}
\includegraphics[width=8cm,height=6cm]{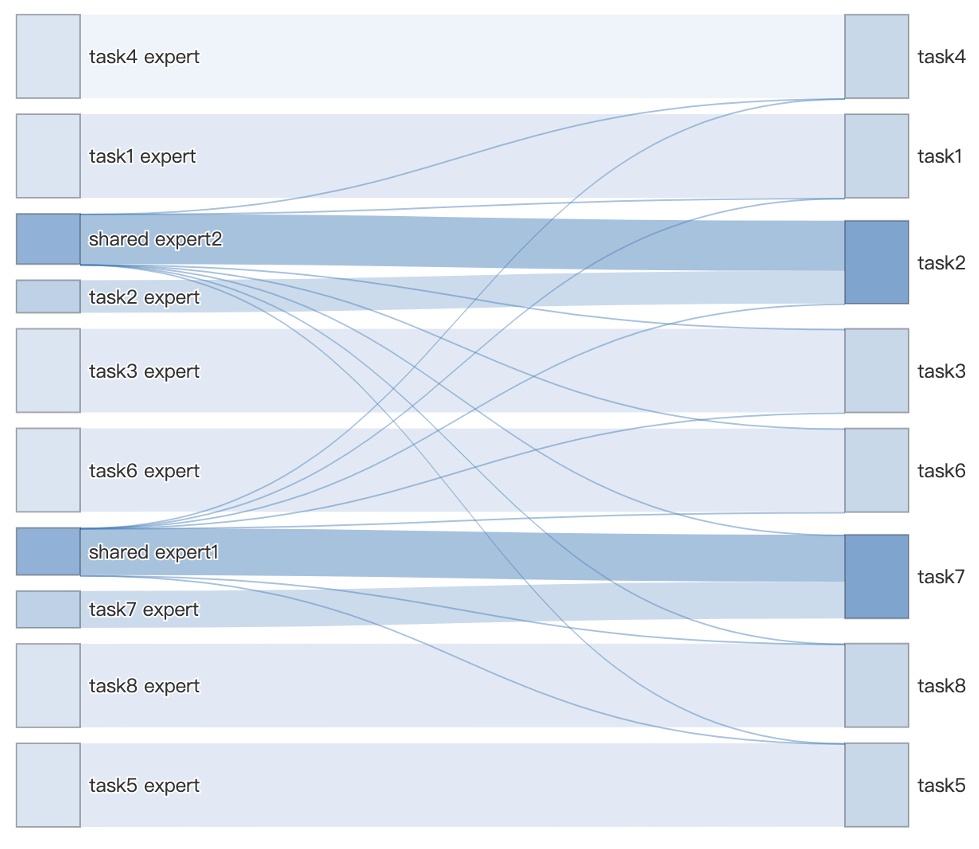}
\caption{Expert degradation issue in CGC, where the two shared experts are almost monopolized by task2 and task7, respectively, working in a specific style.}
\label{expert_degradation_issue}
\end{center}
\end{figure}

    \textbf{Expert Degradation}: 
    \textit{After we fixed the above serious expert collapse issue}, we successfully upgraded our multi-task module to a shared-specific MoE variant, CGC.
    As a result, we are curious whether the gating weights can get the expected results that all task gate networks could assign perceivable scores for shared-experts and their specific-experts to achieve an equilibrium status.
    Unfortunately, we found another unexpected expert degradation phenomenon (as shown in  Figure~\ref{expert_degradation_issue}).
    Here, we show the average scores of the gating mechanisms for some major towers, and we observe that the shared-expert hardly contributes to all tasks but degrades to a specific-expert only belongs to few tasks.
    Therefore, such observation reveals that it is difficult for the architecture of naive shared and specific experts to converge to the ideal status.

\begin{figure}[t]
\begin{center}
\includegraphics[width=8cm,height=6cm]{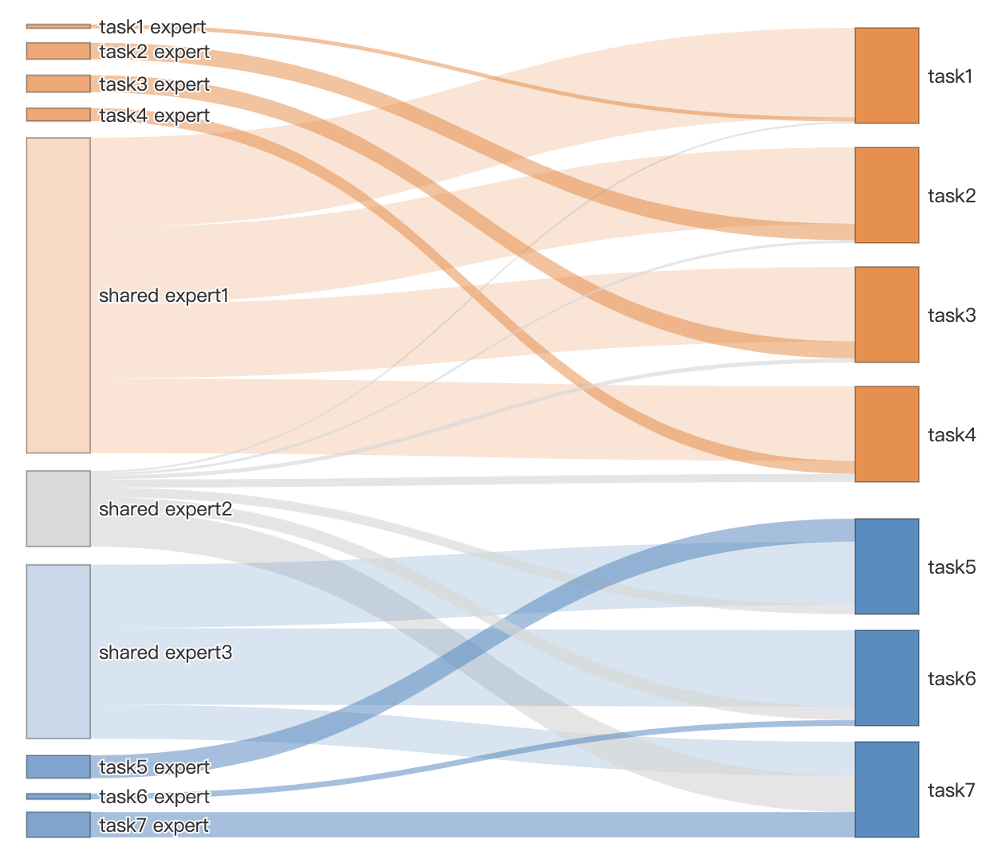}
\caption{Expert underfitting issue, where task1 and task6 almost rely on shared experts only and ignore their own specific expert, making less use of the specific expert network.
}
\label{expert_underfitting_issue}
\end{center}
\end{figure}

    \textbf{Expert Underfitting}: 
    \textit{After we further fixed the expert degradation} and enhanced the efficiency of shared-experts for all tasks, we found some specific-experts are assigned a small gate value so that corresponding tasks only rely on the shared knowledge, making less use of specific parameters.
    Actually, our model needs to predict dozens of different tasks simultaneously, and their densities (i.e., positive sample rate) also vary greatly, while dense tasks can be 100x larger than sparse tasks, e.g., Click v.s. Collect. 
    Compared to shared-experts that could receive multiple gradient updates from multiple dense tasks, specific-experts easily fall into underfitting, further leading the sparse task to rely more on shared-experts but ignoring their specific-experts and making specific parameters wasted.
    As shown in Figure~\ref{expert_underfitting_issue}, the task 6 gate network assigns a large value to the shared-experts but overlooks its specific-experts.

To address these anomalies and improve MoE paradigm model stability, we propose a simple, efficient and balanced neural network architecture for multi-task learning: \underline{H}ierarchy \underline{o}f \underline{M}ulti-gate \underline{E}xperts model, termed as \textbf{HoME}.
Specifically, we provide insightful and in-depth solutions from three perspectives: the value distribution alignment for fair expert weights, the hierarchy meta expert structure to re-assemble tasks, and the gate networks to enhance sparse task expert and deep multi-layer MMoE training:
%

    \textbf{Expert normalization\&Swish mechanism}: To balance the variance of experts outputs and avoid expert collapse, we first introduced the normal~\cite{batchnorm, layernorm} operation for each expert to project their output to approximate the normal distributions, i.e., expert outputs distribution $\thickapprox \mathcal{N}(0, \mathbf{I})$.
    However, under this setting, we found that performing normalization directly will also lead to too many 0 existing after the ReLU function.
    The reason might be that the mean value of normalized expert output is close to 0, thus half of the outputs will be less than 0 and then activated as 0 under ReLU.
    To alleviate the zero derivatives gradient phenomenon, we use the Swish~\cite{swish} function to replace the ReLU function to improve the utilization of parameters and speed up the training process.
    Since the normalization and swish setting, all experts' output could align to a similar numerical magnitude, which could help our gate network assign comparable weights.
    %
    
    \textbf{Hierarchy mask mechanism}: 
    To reduce expert occupancy issues and away from expert degradation (also called task conflict seesaw issue~\cite{ple, star, pepnet}), in this paper, we present a \textit{simple-yet-effective cascading hierarchy mask mechanism} to alleviate such conflict.
    Specifically, we insert a pre-order meta expert network to group different tasks to extend the standardized MoE system.
    %
    As shown in Figure~\ref{kuaishou}, our short-video behaviors tasks could be manually divided into two meta categories according to their prior relevance: (1) passive watching-time tasks, e.g., Long-view; (2) proactive interaction tasks, e.g., Comment.
    Therefore, we can pre-model coarse-grained meta-category experts and then support each task with the following idea: each task should have not only fully-shared global experts, but also partial-shared in-category experts.

    \textbf{Feature-gate and Self-gate mechanisms}: 
    To enhance the training of our sparse-task experts, we present two gate mechanisms to ensure they can obtain appropriate gradients to maximize their effectiveness: feature-gate and self-gate mechanisms.
    Considering that the same layer experts always share the same input features, but different experts will receive different gradients. Thus, the same feature input may lead to the potential risk of gradient conflicts for multiple expert parameter optimization.
    To this end, we first present the feature-gate mechanism to privatize flexible expert inputs to protect sparse-task expert training.
    Besides, the latest MMoE efforts show that deeper stacking expert networks~\cite{ple,adatt} could bring more powerful prediction ability.
    However, in our experiment, we find the origin gate network easily dilutes the gradient layer by layer, which is unfriendly for the sparse-task expert training.
    To ensure the top layers gradient can be effectively passed to the bottom layers and stabilize the deeper MMoE system training, we further devise the self-gate mechanism to connect the adjacent related experts residually.

The main contributions of our work are as follows:
\begin{itemize}[leftmargin=*,align=left]
    \item We deeply analyze the expert issues of the current MoE system and propose our milestone work HoME. To the best of our knowledge, this paper is the first to focus on enhancing multi-task MoE system stability, which will shed light on other researchers to explore a more robust multi-task MoE system.
    \item We conduct extensive offline and online experiments at Kuaishou short-video service. The offline experiments show that all prediction tasks get significant improvements, and the online experiments obtain 0.636\% and 0.735\% play-time improvements on Kuaishou and Kuaishou-Lite applications.
    \item Our HoME has been widely deployed on various services at Kuaishou, supporting 400 million active users daily. 
\end{itemize}

\section{Related Works}
In this section, we briefly review the evolution trajectory of multi-task learning, which plays a more and more important role in empowering models to perceive multiple signals in various research fields, including recommender systems~\cite{rsmtl1, rsmtl2, rsmtl3}, neural language processing~\cite{deepseekmoe, mtlnlp1, mtlnlp2}, computer vision~\cite{cvmtl1, cvmtl2, cvmtl3} and ubiquitous computing~\cite{ubiquitousmtl1, ubiquitousmtl2}.
In early years, several works utilized the hard expert sharing architecture with multi task-specific towers fed by the same expert output to achieve the simplest multi-task learning system, including the shared-bottom~\cite{sharedbottom}, mixture-of-expert (MoE~\cite{moe}).
Later, the cross-stitch~\cite{crossstitch} network and sluice~\cite{sluice} network were proposed to build a deep expert information fusion network to generate the task-specific inputs to achieve soft expert knowledge sharing.
Except the complex vertical deep expert crossing, the horizontal expert weight estimating is another way to customize task-specific tower input, the recently year proposed multi-gate mixture-of-expert (MMoE)~\cite{mmoe} gives a multi-gate mechanism to assign different weights to different experts in order to balance different tasks.
With the wave of neural networks-based recommender systems, the MMoE variants methods also play a significant role in improving the model capabilities and accuracy.
The pioneering work is from the Youtube ranking system~\cite{youtube2}, which utilizes several shared experts through different gating networks to model the real user-item interactions.
To alleviate the task-conflict seesaw~\cite{ple, star, pepnet} problem, the MMoE variants CGC~\cite{ple} and PLE~\cite{ple} not only utilize shared-experts, but also insert additional specific-experts for more flexible expert sharing.
Based on the shared/specific idea, a lot of MMoE variant was proposed, including:
MSSM~\cite{mssm} extends the PLE approach by employing a field-level and cell-level features selective mechanism to determine the importance of input features automatically.
AdaTT~\cite{adatt} leveraging an adaptive fusion gate mechanism on PLE to model complex task relationships between specific-expert and shared-expert.
STAR~\cite{star} adopts star topology with one shared expert network and some specific networks to fuse expert parameters.
MoLA~\cite{MoLA} borrows the low-rank fine-tuning technique from LLM and devices lightweight low-rank specific-expert adapters to replace complex specific-expert.

\begin{figure*}[t]
\begin{center}
\includegraphics[width=18cm,height=6cm]{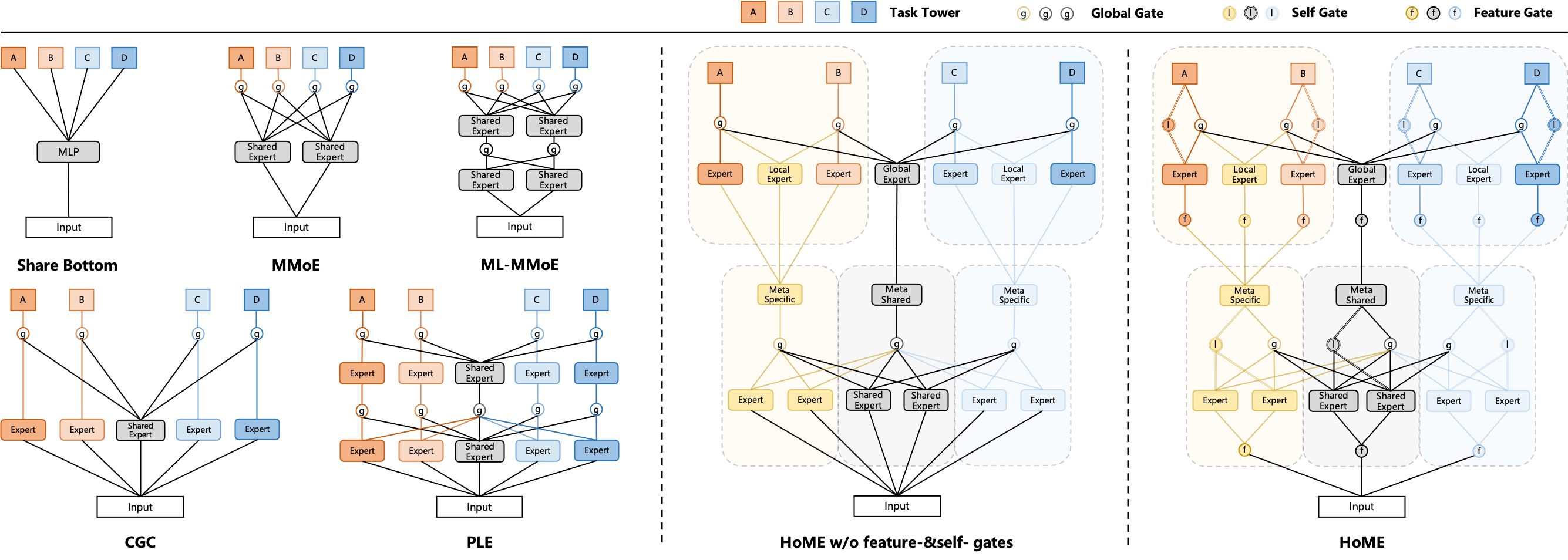}
\caption{The HoME and other MoE-style multi-task learning architectures. In HoME, tasks are divided into groups based on their relatedness and modeled as fully-shared or partial-shared meta-representations in the first layer, then refined as specific task representations in the second layer. HoME further introduces two specially designed modules: Feature-gate to alleviate task conflicts at the input level, and Self-gate to ensure that each task makes the most of specific experts. Best viewed in color.}
\label{related}
\end{center}
\end{figure*}

\section{Methodology}
In this section, we introduce the components of our model, HoME.
We first retrospect how the MoE system works in an industrial scale RecSys, from feature engineering, MoE neural networks details and prediction scores assembled for ranking.
Afterward, we express our solution for the three problems: expert normalization\&swish mechanism to overcome the expert collapse issue,  hierarchy mask mechanism to alleviate expert degradation issue, and two kinds of gate mechanisms for the expert underfitting issue.

\subsection{Preliminary: Multi-Task Learning for Industrial Recommender System}
The industrial recommender system follows a two-stage design: (1) hundreds of item candidate generation~\cite{hstu, yan2024trinity} and (2) item candidate ranking~\cite{essm, youtube, youtube2} to select dozens of top items for users.
Since the goals of these two stages are distinct, thus the techniques used are also completely different: the generation process focuses on user-side feature modeling and coarsen item sampling while the ranking process focuses on user and item feature fusion and fine-grained user multi-interaction fitting. 
Therefore, the multi-task learning model is always employed in the ranking process, to estimate various interactions' probabilities for a specific user-item pair.
For brevity, the model-generated probabilities always have a \textbf{short name (xtr)}, e.g., click probability as \textbf{ctr}, effective-view probability as \textbf{evtr}, like probability as \textbf{ltr}, comment probability as \textbf{cmtr}, and so on.

\subsubsection{Label\&Feature}
Formally, such a ranking learning process is always organized as a multiple binary classifications style, and each learning user-item samples contain two types of information -- the supervised label and input features: 
\begin{itemize}[leftmargin=*,align=left]
    \item \textit{Supervised Signals}: the real labels of this user-item watch experience, e.g., click $y^{ctr}\in\{0,1\}$, effective view $y^{evtr}\in\{0,1\}$, like $y^{ltr}\in\{0,1\}$, comment $y^{cmtr}\in\{0,1\}$ and other labels.
    \item \textit{Feature Inputs}: the \textbf{MoE input} aims to describe the status of user and item from multiple perspectives and can be roughly divided into four classes: (1) ID and category features, we use a straightforward lookup operator to get their embeddings, e.g., user ID, item ID, tag ID, is active user, is follow author, Scenario ID and others; (2) statistics features, which needs to devise bucketing strategies to discretize them to assign an ID for them, e.g., number of watched short-video in the last month, short-video viewing time in the past month, and others. (3) the sequential features to reflect users short-term and long-term interests, which usually is modeled by the one-stage or two-stage attention mechanism, e.g., DIN~\cite{din}, DIEN~\cite{dien}, SIM~\cite{sim}, TWIN~\cite{twin}. (4) pre-trained multi-modal embeddings such as text embedding~\cite{bert}, asr embedding~\cite{speechlm}, video embedding~\cite{liu2024sora}, etc.
\end{itemize}
Combination all of them, we can obtain the multi-task training samples (e.g., labels are $\{y^{ctr}, y^{evtr}, \dots\}$, inputs are $\mathbf{v} = [\mathbf{v}_1, \mathbf{v}_2, \dots, \mathbf{v}_n]$), where $n$ indicates the total feature number.

\subsubsection{Mixture-of-Experts for XTR prediction}
Given the training user-item sample labels $y^{ctr}, y^{evtr}, \dots$ and features $\mathbf{v}$, next we utilize the multi-task module to make predictions.
Specifically, we show the wide-used shared/specific paradigm MoE variant CGC details as follows:
%
\begin{equation}
\small
\begin{split}
\hat{y}^{ctr} &= \texttt{Tower}^{ctr}\big(\texttt{Sum}\big(\texttt{Gate}^{ctr}(\mathbf{v}), \{\texttt{Experts}^{\{shared, ctr\}}(\mathbf{v})\}\big)\big),\\
\hat{y}^{evtr} &= \texttt{Tower}^{evtr}\big(\texttt{Sum}\big(\texttt{Gate}^{evtr}(\mathbf{v}), \{\texttt{Experts}^{\{shared, evtr\}}(\mathbf{v})\}\big)\big),\\
\hat{y}^{ltr} &= \texttt{Tower}^{ltr}\big(\texttt{Sum}\big(\texttt{Gate}^{ltr}(\mathbf{v}), \{\texttt{Experts}^{\{shared, ltr\}}(\mathbf{v})\}\big)\big),\\
\texttt{whe}&\texttt{re}\quad\quad\quad \texttt{Tower}(\cdot) = \texttt{Sigmoid}\big(\texttt{MLP\_T}(\cdot)\big), \\
& \quad\quad\quad\quad \texttt{Experts}(\cdot) = \texttt{ReLU}\big(\texttt{MLP\_E}(\cdot)\big), \\
& \quad\quad\quad\quad \texttt{Gate}(\cdot) = \texttt{Softmax}\big(\texttt{MLP\_G}(\cdot)\big), \\
\end{split}
\label{cgc}
\end{equation}
where the $\texttt{Expert}^{shared}:\mathbb{R}^{|\mathbf{v}|}\to \mathbb{R}^{D}$ and $\texttt{Expert}^{xtr}:\mathbb{R}^{|\mathbf{v}|}\to \mathbb{R}^{D}$ are the \textbf{ReLU}-activated shared and specific experts networks respectively, the $\texttt{Gate}^{xtr}:\mathbb{R}^{|\mathbf{v}|}\to \mathbb{R}^{N}$ is the \textbf{Softmax}-activated gate network for corresponding task, $N$ is the related shared and specific experts number, $\texttt{Sum}$ aims to aggregate the $N$ experts outputs according to gate-generated weights, $\texttt{Tower}^{xtr}: \mathbb{R}^{D}\to \mathbb{R}$ is the \textbf{Sigmoid}-activated task-specific network to measure corresponding interaction probability $\hat{y}$.

After obtaining all the estimated scores $\hat{y}^{ctr}, \dots$ and ground-truth labels $y^{ctr}, \dots$, we directly minimize the cross-entropy binary classification loss to train the multi-task learning model:
\begin{equation}
\small
\begin{split}
\mathcal{L} = - \sum_{{ctr, \dots}}^{xtr} \big(y^{xtr}\log{(\hat{y}^{xtr})} + (1-y^{xtr})\log{(1-\hat{y}^{xtr}})\big)
\end{split}
\label{crossentropy}
\end{equation}
In online serving, a common operation is to devise a controllable complex equation to combine XTRs as one ranking score:
\begin{equation}
\small
\begin{split}
\texttt{ranking\_score} = \alpha\cdot\hat{y}^{ctr} + \beta\cdot\hat{y}^{evtr} + \gamma\cdot\hat{y}^{cmtr} + \dots
\end{split}
\label{rankingscore}
\end{equation}
where $\alpha$, $\beta$, $\gamma$ are hyper-parameters.
In fact, Eq.(\ref{rankingscore}) is very complicated with many strategies in industry RecSys. We only show a naive case.
In the following section, we focus on the multi-task learning procedure in Eq.(\ref{cgc}) to improve its stability.

\subsection{Expert Normalization\&Swish Mechanism}
Although the vanilla MMoE system in Eq.(\ref{cgc}) achieves remarkable improvements, it still exists the serious expert collapse problem.
Denote the experts' $\texttt{MLP\_E}$ function generated representation as \{$\mathbf{z}^{shared}$, $\mathbf{z}^{ctr}$, $\mathbf{z}^{evtr}$,\dots\}, we found their means and variances values are significantly different.
Inspired by the Transformers, the normalization operator is one of the vital techniques to successfully support training very deep neural networks.
We also introduce the batch normalization~\cite{batchnorm} for each expert to support our HoME to generate comparable output $\mathbf{z}_{norm}\in \mathbb{R}^D$:
\begin{equation}
\small
\begin{split}
\mathbf{z}&_{norm} = \texttt{Batch\_Normalization}(\mathbf{z}) = \bm{\gamma}\frac{\mathbf{z} - \bm{\mu}}{\sqrt{\bm{\delta}^2 + \bm{\epsilon}}} + \bm{\beta},\\
\texttt{Where} &\quad \bm{\mu} = \texttt{Batch\_Mean}(\mathbf{z}), \quad \bm{\delta}^2 = \texttt{Batch\_Mean}\big((\mathbf{z} - \bm{\mu})^2\big), \\
\end{split}
\label{batchnorm}
\end{equation}
where $\mathbf{z}$ is an arbitrary experts' $\texttt{MLP\_E}$ output, $\bm{\gamma} \in \mathbb{R}^D$, $\bm{\mu} \in \mathbb{R}^D$ are trainable scale and bias parameters to adjust the distribution, $\bm{\epsilon} \in \mathbb{R}^D$ is a very small factor to avoid the division by 0 error.
$\bm{\mu} \in \mathbb{R}^D$, $\bm{\delta}^2 \in \mathbb{R}^D$ are mean and variances of \textbf{current batch same expert outputs}.
After the expert normalization, the distribution of $\mathbf{z}_{norm}$ is a normal distribution that is closely related to $\mathcal{N}(0, \mathbf{I})$.
As a result, the half of $\mathbf{z}_{norm}$ values will be less than 0 and then activated as 0 under ReLU, causing their derivatives and gradients to be 0, cumbering model convergence.
Thus, we use the Swish function to replace the ReLU in Eq.(\ref{cgc}) to obtain our HoME Expert:
\begin{equation}
\footnotesize
\begin{split}
\texttt{HoME\_Expert}(\cdot) = \texttt{Swish}\Big(\texttt{Batch\_Normalization}\big(\texttt{MLP\_E}(\cdot)\big)\Big),
\end{split}
\label{homeexpert}
\end{equation}
where the $\texttt{HoME\_Expert}(\cdot)$ is the final structures used in our HoME.
Under the normalization and swish setting, the output of all experts could align to a similar numerical magnitude, which could help our gate network assign comparable weights.
For brevity, in the following section, \textbf{we still use} $\texttt{Expert}(\cdot)$ \textbf{to represent our} $\texttt{HoME\_Expert}(\cdot)$.

\subsection{Hierarchy Mask Mechanism}
For the expert degradation, there is a series of works that introduce novel specific-expert and shared-expert architecture to alleviate task conflicts.
However, following the specific and shared paradigm, we found that the problem of shared expert degradation still occurs.
We argue that it can be beneficial to consider the prior task relevance, as shown in Figure~\ref{kuaishou}; our prediction task can be divided into two categories, e.g., proactive interaction tasks (e.g., Like, Comment, etc.) and passive watching-time tasks (e.g., Effective-view, Long-view, etc.).
In this section, we propose a simple-yet-effective cascading hierarchy mask mechanism to model the prior inductive bias between tasks.
Specifically, we insert a pre-order meta expert network to group different tasks, here including three meta-task knowledge to support our two categories of tasks:
\begin{equation}
\footnotesize
\begin{split}
\mathbf{z}^{inter}_{meta}& = \texttt{Sum}\big(\texttt{Gate}^{inter}_{meta}(\mathbf{v}), \{\texttt{Experts}^{\{shared, inter\}}_{meta}(\mathbf{v})\}\big),\\
\mathbf{z}^{watch}_{meta}& = \texttt{Sum}\big(\texttt{Gate}^{watch}_{meta}(\mathbf{v}), \{\texttt{Experts}^{\{shared, watch\}}_{meta}(\mathbf{v})\}\big),\\
\mathbf{z}^{shared}_{meta} =& \texttt{Sum}\big(\texttt{Gate}^{shared}_{meta}(\mathbf{v}), \{\texttt{Experts}^{\{shared, inter, watch\}}_{meta}(\mathbf{v})\}\big),\\
\end{split}
\label{home_1}
\end{equation}
where $\mathbf{z}^{inter}_{meta}$, $\mathbf{z}^{watch}_{meta}$, $\mathbf{z}^{shared}_{meta}$ are coarsen macro-level meta representation to extract: (1) interaction in-category knowledge, (2) watch-time in-category knowledge and (3) shared knowledge.

After obtaining these meta representations, we next focus on the multi-task prediction according to their corresponding meta knowledge and shared meta knowledge.
Specifically, we utilize the meta knowledge to build three types of experts: (1) the globally shared experts for all tasks according to $\mathbf{z}^{shared}_{meta}$, (2) the locally shared experts for in-category tasks according to $\mathbf{z}^{inter}_{meta}$ or $\mathbf{z}^{watch}_{meta}$, (3) the specific experts for each task according to $\mathbf{z}^{inter}_{meta}$ or $\mathbf{z}^{watch}_{meta}$.
For the task-specific gate networks, we directly use the concatenation shared meta knowledge $\mathbf{z}^{shared}_{meta}$ and corresponding category meta knowledge to generate the weights of experts.
Here, we take the Click and Effective-view interactions as examples:
\begin{equation}
\footnotesize
\begin{split}
\hat{y}^{ctr} =  \texttt{Tower}^{ctr}(\texttt{Sum}&\big(\texttt{Gate}^{ctr}(\mathbf{z}^{inter}_{meta}\oplus\mathbf{z}^{shared}_{meta}), \\ 
&\{\texttt{Experts}^{shared}(\mathbf{z}^{shared}_{meta}), \\
& \texttt{Experts}^{\{inter, ctr\}}(\mathbf{z}^{inter}_{meta})\}\big),\\
\hat{y}^{evtr} =  \texttt{Tower}^{evtr}(\texttt{Sum}&\big(\texttt{Gate}^{evtr}(\mathbf{z}^{watch}_{meta}\oplus\mathbf{z}^{shared}_{meta}), \\ 
&\{\texttt{Experts}^{shared}(\mathbf{z}^{shared}_{meta}), \\
& \texttt{Experts}^{\{watch, evtr\}}(\mathbf{z}^{watch}_{meta})\}\big),\\
\end{split}
\label{home_2}
\end{equation}
where the $\oplus$ denotes the concatenation operator, the $\texttt{Experts}^{shared}$ are the all tasks shared experts, $\texttt{Experts}^{inter}$, $\texttt{Experts}^{watch}$ are the in-category tasks shared experts.

\begin{table*}[t!]
\centering
\caption{Offline results (\%) (AUC and GAUC) on Short-Video services at Kuaishou.}
\resizebox{\linewidth}{!}{
\begin{tabular}{l|cccccccccccccccc|c}
\toprule
\multirow{2}{*}{\textbf{Model}} & \multicolumn{2}{c}{\textbf{Effective-view}} & \multicolumn{2}{c}{\textbf{Long-view}} & \multicolumn{2}{c}{\textbf{Click}} & \multicolumn{2}{c}{\textbf{Like}} & \multicolumn{2}{c}{\textbf{Comment}} & \multicolumn{2}{c}{\textbf{Collect}} & \multicolumn{2}{c}{\textbf{Forward}} & \multicolumn{2}{c|}{\textbf{Follow}} & \multirow{2}{*}{\textbf{\#Params}}         
\\  & AUC & GAUC & AUC & GAUC & AUC & GAUC & AUC & GAUC & AUC & GAUC & AUC & GAUC & AUC & GAUC & AUC & GAUC &\\
\midrule
MMoE & 77.56 & 71.90 & 82.91 & 77.04 & 73.43 & 69.38 & 96.94 & 84.85 & 92.44 & 78.55 & 92.85 & 80.12 & 92.44 & 76.75 & 95.42 & 84.30 & 224.70Mil \\
\midrule
MMoE* & 77.66 & 72.03 & 82.98 & 77.15 & 73.66 & 69.62 & 96.96 & 84.97 & 92.49 & 78.68 & 92.91 & 80.27 & 92.53 & 76.93 & 95.50 & 84.51 & 224.85Mil \\
CGC* w/o shared & 77.72 & 72.10 & 83.03 & 77.21 & 73.84 & 69.85 & \underline{97.02} & 85.11 & 92.51 & 78.76 & 93.03 & 80.51 & 92.67 & 77.12 & 95.57 & 84.70 & 279.43Mil \\
CGC* & 77.72 & 72.11 & 83.04 & 77.23 & 73.88 & 69.89 & \underline{97.02} & 85.12 & 92.51 & 78.78 & 93.03 & 80.53 & 92.68 & 77.16 & 95.59 & 84.76 & 325.83Mil  \\
PLE* & 77.74 & 72.14 & 83.04 & 77.24 & \underline{73.92} & \underline{69.92} & \underline{97.02} & \underline{85.15} & \underline{92.54} & \underline{78.82} & \underline{93.05} & \underline{80.57} & \underline{92.70} & \underline{77.22} & \underline{95.61} & \underline{84.80} & 351.29Mil \\
AdaTT* & \underline{77.76} & \underline{72.16} & \underline{83.07} & \underline{77.27} & 73.84 & 69.83 & 97.01 & 85.12 & 92.53 & 78.79 & 92.98 & 80.45 & \underline{92.70} & 77.18 & 95.59 & 84.73 & 305.01Mil \\
HoME & \textbf{77.87} & \textbf{72.34} & \textbf{83.19} & \textbf{77.42} & \textbf{73.95} & \textbf{69.98} & \textbf{97.03} & \textbf{85.23} & \textbf{92.61} & \textbf{79.03} & \textbf{93.12} & \textbf{80.77} & \textbf{92.76} & \textbf{77.42} & \textbf{95.64} & \textbf{84.87} & 292.24Mil \\
\midrule
\textbf{Improve over MMoE} & +0.31 & +0.44 & +0.28 & +0.38 & +0.52 & +0.60 & +0.09 & +0.38 & +0.17 & +0.48 & +0.27 & +0.65 & +0.32 & +0.67 & +0.22 & +0.57 & - \\
\midrule
\midrule
HoME w/o fg2 & 77.85 & 72.30 & 83.16 & 77.39 & 73.94 & 69.95 & 97.02 & 85.19 & 92.60 & 78.99 & 93.10 & 80.71 & 92.74 & 77.34 & 95.62 & 84.83 & 268.18Mil \\
HoME w/o fg & 77.78 & 72.22 & 83.11 & 77.32 & 73.89 & 69.89 & 97.01 & 85.15 & 92.58 & 78.91 & 93.06 & 80.62 & 92.70 & 77.24 & 95.60 & 84.77 & 204.51Mil\\
HoME w/o fg-sg & 77.77 & 72.19 & 83.09 & 77.29 & 73.83 & 69.83 & 97.02 & 85.14 & 92.58 & 78.90 & 93.05 & 80.60 & 92.70 & 77.22 & 95.60 & 84.75 & 202.38Mil \\
HoME w/o fg-sg-mask & 77.63 & 72.01 & 82.96 & 77.13 & 73.68 & 69.65 & 96.98 & 84.98 & 92.47 & 78.61 & 92.95 & 80.35 & 92.54 & 76.97 & 95.51 & 84.52 & 202.70Mil \\
\bottomrule
\end{tabular}
}
\small
For a fair comparison, baselines remarked with `*' are equipped with our \texttt{HoME\_Expert} as base Expert network. CGC* w/o shared removes shared experts and all gate networks of CGC*. For HoME, the `w/o fg2' and `w/o fg' variants ignore the second layer feature-gates and all feature-gates respectively; the `w/o sg' variant ignores all self-gates, the `w/o mask' variant keeps HoME architecture but all experts are shared. Best/runner-up results are marked \textbf{bold}/\underline{underlined}.
\label{mainoffline}
\end{table*}

It is worth noting the meta abstraction of HoME's first layer, the main architecture difference with PLE, which is based on our observation of real multi-task recommendation scenario at Kuaishou (see Figure~\ref{related}). Based on the prior semantics divided meta expert network of our HoME, we can avoid conflicts between tasks as much as possible and maximize the sharing efficiency among tasks.

\subsection{Feature-gate\&Self-gate mechanisms}
For the expert underfitting, we find some data-sparse tasks' gate-generated weights tend to ignore their specific experts but assign large gate weights to shared experts.
The reason might be that our model needs to predict 20+ different tasks simultaneously, but these dense tasks' density can be 100x larger than sparse tasks.
To enhance our sparse task expert training, we present two gate mechanisms to ensure they can obtain appropriate gradients to maximize their effectiveness: the feature-gate and self-gate mechanisms.

For feature-gate, the purpose is to generate different representations of input features for different task experts, to alleviate the potential gradient conflicts when all experts share the same input features.
Formally, the feature-gate aims to extract the importance of each input feature element, e.g., $\texttt{Fea\_Gate}:\mathbb{R}^{|\mathbf{v}|}\to \mathbb{R}^{|\mathbf{v}|}$ if the input is $\mathbf{v}$.
However, in industrial recommender systems, the $\mathbf{v}$ is always a high-dimension vector, e.g., $|\mathbf{v}| > 3000+$; thereby, it is expensive to introduce these large matrices for meta experts.
Inspired by the LLM efficiency tuning technique, LoRA~\cite{lora}, we also introduce two small matrices to approximate a large matrix to generate element importance:
\begin{equation}
\small
\begin{split}
&\texttt{Fea\_LoRA}(\mathbf{v}, d) = 2\times\texttt{Sigmoid}\big(\mathbf{v}(\mathbf{B}\mathbf{A})\big), \\
\texttt{where}& \quad \mathbf{B}\in \mathbb{R}^{|\mathbf{v}|\times d},\ \  \mathbf{A}\in \mathbb{R}^{d\times|\mathbf{v}|},\ \  \mathbf{B}\mathbf{A}\in \mathbb{R}^{|\mathbf{v}|\times |\mathbf{v}|}\\
\end{split}
\label{feature_gate}
\end{equation}
Note that we apply a $2\times$ operator after the \texttt{Sigmoid} function, which aims to achieve a flexible zoom-in or zoom-out operator.
Indeed, the $\texttt{Fea\_LoRA}$ function is an effective way to generate privatized expert inputs.
In our iteration, we find it could be further enhanced with multi-task idea, i.e., introducing more $\texttt{Fea\_LoRA}$ to generate feature importance from multiple aspects as our $\texttt{Fea\_Gate}$.
\begin{equation}
\small
\begin{split}
\texttt{Fea\_Gate}(\mathbf{v}) = \texttt{Sum}\big(\texttt{Gate}^{fea}(\mathbf{v}), \{\texttt{Fea\_LoRA}^{\{1, 2, \dots, L\}}(\mathbf{v}, |\mathbf{v}|/L)\}\big),
\end{split}
\label{feature_gate2}
\end{equation}
where $L$ is a hyper-parameter to control the $\texttt{Fea\_LoRA}$ number, the $\texttt{Gate}^{fea}:\mathbb{R}^{|\mathbf{v}|}\to \mathbb{R}^{L}$ utilized to generate weights to balance different $\texttt{Fea\_LoRA}$ importance.
Note that we need to choose an $L$ that is divisible by input length $|\mathbf{v}|$ to generate the dimension of \texttt{Fea\_LoRA}.
Therefore, our expert input can be obtained as follows (here we show the first layer meta shared experts input $\mathbf{v}^{shared}_{meta}$):
\begin{equation}
\small
\begin{split}
\mathbf{v}^{shared}_{meta} = \mathbf{v}\odot \texttt{Fea\_Gate}^{shared}_{meta}(\mathbf{v}),
\end{split}
\label{feature_gate3}
\end{equation}
where the $\odot$ denotes the element-wise product.
In this way, the different experts have their own feature space, which could reduce the risk of gradient conflicts to protect sparse tasks.

Besides, the latest MoE efforts show that deeper stacking expert networks could bring more powerful prediction ability~\cite{ple,adatt}.
Unfortunately, in our experiment, we find the origin gate network easily dilutes the gradient layer by layer, especially for the sparse task expert training.
In addition to the expert-input level $\texttt{Fea\_Gate}$, we also add a residual idea-based self-gate on the expert-output level to ensure the top layers gradient can be effectively passed to bottom layers.
Specifically, the $\texttt{Self\_Gate}$ only focuses on the output of its specific experts. Take the watching-time meta experts output as an example:
\begin{equation}
\small
\begin{split}
&\mathbf{z}^{shared}_{meta,self}=\texttt{Sum}\Big(\texttt{Self\_Gate}^{shared}_{meta}(\mathbf{v}), \{\texttt{Experts}^{shared}(\mathbf{v})\}\Big)\\
&\texttt{Where} \ \  \texttt{Self\_Gate}(\cdot) = \texttt{Sigmoid}\big(\texttt{MLP\_G}(\cdot)\big) \ \  \texttt{if only 1 Expert} \\
&\quad\quad \ \ \ \ \texttt{Self\_Gate}(\cdot) = \texttt{Softmax}\big(\texttt{MLP\_G}(\cdot)\big) \ \  \texttt{others}
\end{split}
\label{feature_gate3}
\end{equation}
where $\texttt{Self\_Gate}:\mathbb{R}^{|\mathbf{v}|}\to \mathbb{R}^{K}$, where $K$ is the related \texttt{Expert} number, and its activate function is \texttt{Sigmoid} if there only 1 \texttt{Expert}, otherwise setted as \texttt{Softmax}.
Analogously, the $\mathbf{z}^{inter}_{meta,self}$, $\mathbf{z}^{watch}_{meta,self}$ can be obtained in the same way, we then add the corresponding representations (e.g., $\mathbf{z}^{inter}_{meta}$ + $\mathbf{z}^{inter}_{meta,self}$) to support the next layer.
See Figure~\ref{related} for fine-grained HoME details.

\section{Experiments}
In this section, we first compare HoME with several widely-used multi-task learning approaches in offline settings.
We then conducted some model variations with our modifications to verify the effectiveness of HoME.
We also test the impact of HoME hyper-parameters robustness on the number of expert numbers and feature-gate LoRA numbers. 
Furthermore, we provide our model's expert network gate weights graph to show our HoME is promising to be a balanced system.
Finally, we push our HoME to the online A/B test to verify how much benefit that HoME can contribute to Kuaishou.

\subsection{Experiments Setup}
We conduct experiments at our short-video data-streaming, which is the largest recommendation scenario at Kuaishou, including over \textbf{400 Million users} and \textbf{50 Billion logs} every day.
For a fair comparison, we only change the multi-task learning module in Eq.(\ref{cgc}), and keep the same of other modules.
Specifically, we implement the MMoE~\cite{mmoe}, CGC~\cite{ple}, PLE~\cite{ple}, AdaTT~\cite{adatt} model variants as baselines.
For the evaluation, we use the wide-used ranking metrics AUC and GAUC~\cite{din} to reflect the model's predictive ability.
Specifically, in our short-video service, GAUC is the most important offline metric. Its main idea is to calculate each user's AUC separately and then weighted aggregate all users' AUC as:
\begin{equation}
\small
\begin{split}
\texttt{GAUC} = \sum_{u} w_u \texttt{AUC}_u \quad \texttt{where}\ \ w_u = \frac{\texttt{\#logs}_u}{\sum_i \texttt{\#logs}_i},
\end{split}
\label{gauc}
\end{equation}
where the $w_u$ denotes the user's logs ratio.

\subsection{Offline Experiments}
The main experiment results are shown in Table~\ref{mainoffline}. Note that improvements of \textbf{0.03\%\textasciitilde0.05\%} in AUC or GAUC in the offline evaluation are significant enough to bring substantial online revenue to our business.
We first show the effectiveness of the \texttt{HoME\_Expert} upon MMoE, i.e., MMoE*. Then we compare HoME with the improved baselines all equipped with \texttt{HoME\_Expert}, such as `CGC* w/o shared', the variant of CGC ignores the shared experts and all gate networks.
Moreover, we also implement ablation variants for our HoME; the `w/o fg2' and `w/o fg' variants ignore the second layer feature-gates and all feature-gates respectively; the `w/o sg' variant ignores all self-gates, the `w/o mask' variant keeps HoME architecture but all experts are shared.
We have the following observations:
\begin{itemize}[leftmargin=*,align=left]
\item 
(1) The MMoE* largely outperforms the naive MMoE method, which indicates our Expert normalization\&Swish mechanism could overcome the expert collapse issue, balance expert outputs and encourage the expert networks to take due responsibility.
(2) The `CGC* w/o shared' can be seen as Shared-bottom with a specific-expert for each task. The MMoE* is weaker than the trivial `CGC* w/o shared' solution equipped with more parameters (24\% bigger compared to MMoE* in our experiment), which indicates that MMoE systems are fragile and can easily degrade in real large-scale streaming data scenarios.
(3) Compared to `CGC* w/o shared', the CGC* does not show significant improvement, which indicates the shared-experts of CGC* are degenerating into some specific-experts.
(4) Compared to MMoE*, the PLE* and AdaTT* achieve better performance, which indicates that after solving expert collapse, stacking multiple expert network layers and increasing model parameters are a promising way to unleash the potential of multi-task modules.
(5) HoME shows statistical improvements over other strong baselines in all tasks, while introducing fewer parameters and achieving the best results, which indicates our modification could enhance multi-task MoE system stability and maximize expert efficiency.
%
%
%
%
%
%

\item (1) For our HoME ablations, the `w/o fg-sg-mask' variant shows comparable performance with MMoE*, while the `/o fg-sg' variant achieves significant improvement across all tasks, i.e., AUC +0.15\% in most cases, which demonstrates that our \textbf{Hierarchy Mask Mechanism is a powerful and low-resources strategy} to alleviate expert degradation issue without introduce large additional parameters.
(2) The `w/o fg' variant reaches better and more steady improvements than the `w/o fg-sg' variant, which indicates adding the residual connection between different layer experts is helpful to train experts.
(3) Compared with HoME and the `w/o fg2' variant, we can find the second layer feature-gates could enhance model ability, but the first layer feature-gates show more robust and greater improvements. The reason might be that the first layer is the input as the information source and used in the coarsen meta layer; their tasks gradient conflict problem will be more serious than the second fine-grained layer.

%
%
%
%
\end{itemize}

\subsection{Discussion of Hyper-Parameter Sensitivity}
This section explores the hyper-parameter sensitivity of the expert's numbers and the feature-gate LoRA numbers, to investigate the robustness of HoME.
For the expert number, we conduct experiments under the `HoME w/o fg' variant, since the first layer feature-gate is an expensive parameter-consuming operator.
From Table~\ref{ablationstudyexpert}, we can observe a HoME scaling-law phenomenon: 
only by introducing more experts, the prediction accuracy will steadily improve with the increase in the number of parameters.
Such a phenomenon also demonstrates our HoME is a balanced MoE system, which could unleash the ability of all experts.
For the feature LoRA number, we conduct experiments under the variant `HoME w/o fg2', which only involves the first layer feature-gate while showing a significant improvements in Table~\ref{mainoffline}.
Specifically, in our implementation, more LoRA numbers will only reduce the dimension of the hidden dimension while not adding additional parameters, which may decrease single LoRA ability.
From Table~\ref{ablationstudylora}, we can observe that the variant of two LoRA shows the best results, which indicates there exists a bottleneck to balance the LoRA number and LoRA modeling ability to provide more incremental information.
%
%
%
%
%


\begin{table*}[t]
\centering
\caption{
Hyper-Parameter Sensitivity discussion of expert networks regarding the number of expert numbers.}
\vspace{-1mm}
\resizebox{\linewidth}{!}{
\begin{tabular}{l|c|cccccccccccccccc|c}
\toprule
\multirow{2}{*}{\textbf{Variant}} & \multirow{2}{*}{\makecell{\textbf{Expert}\\\textbf{Number}}} & \multicolumn{2}{c}{\textbf{Effective-view}} & \multicolumn{2}{c}{\textbf{Long-view}} & \multicolumn{2}{c}{\textbf{Click}} & \multicolumn{2}{c}{\textbf{Like}} & \multicolumn{2}{c}{\textbf{Comment}} & \multicolumn{2}{c}{\textbf{Collect}} & \multicolumn{2}{c}{\textbf{Forward}} & \multicolumn{2}{c|}{\textbf{Follow}} & \multirow{2}{*}{\textbf{\#Parameter}} 
\\  & & AUC & GAUC & AUC & GAUC & AUC & GAUC & AUC & GAUC & AUC & GAUC & AUC & GAUC & AUC & GAUC & AUC & GAUC\\
\midrule
\multirow{4}{*}{\makecell{HoME w/o fg}} & 1 & 77.78 & 72.22 & 83.11 & 77.32 & 73.89 & 69.89 & 97.01 & 85.15 & 92.58 & 78.91 & 93.06 & 80.62 & 92.70 & 77.24 & 95.60 & 84.77 & 204.51Mil     \\
\multirow{4}{*}{} & 2 & 77.81 & 72.26 & 83.13 & 77.36 & 73.90 & 69.92 & 97.02 & 85.18 & 92.59 & 78.94 & 93.08 & 80.68 & 92.72 & 77.28 & 95.62 & 84.80 & 243.28Mil      \\
\multirow{4}{*}{} & 3 & 77.83 & 72.29 & 83.15 & 77.37 & 73.94 & 69.93 & \textbf{97.03} & 85.19 & 92.60 & 78.97 & 93.10 & 80.73 & 92.74 & 77.33 & 95.63 & 84.85 & 282.04Mil      \\
\multirow{4}{*}{} & 4 & \textbf{77.85} & \textbf{72.31} & \textbf{83.17} & \textbf{77.40} & \textbf{73.96} & \textbf{69.95} & \textbf{97.03} & \textbf{85.21} & \textbf{92.61} & \textbf{78.99} & \textbf{93.11} & \textbf{80.77} & \textbf{92.76} & \textbf{77.38} & \textbf{95.66} & \textbf{84.89} & 320.81Mil     \\

\bottomrule
\end{tabular}
}
\label{ablationstudyexpert}
\end{table*}

\begin{table*}[t]
\centering
\caption{Hyper-Parameter Sensitivity discussion of Feature-Gate regarding the number of LoRA.}
\vspace{-1mm}
\resizebox{\linewidth}{!}{
\begin{tabular}{l|c|cccccccccccccccc|c}
\toprule
\multirow{2}{*}{\textbf{Variant}} & \multirow{2}{*}{\makecell{\textbf{LoRA}\\\textbf{Number}}} & \multicolumn{2}{c}{\textbf{Effective-view}} & \multicolumn{2}{c}{\textbf{Long-view}} & \multicolumn{2}{c}{\textbf{Click}} & \multicolumn{2}{c}{\textbf{Like}} & \multicolumn{2}{c}{\textbf{Comment}} & \multicolumn{2}{c}{\textbf{Collect}} & \multicolumn{2}{c}{\textbf{Forward}} & \multicolumn{2}{c|}{\textbf{Follow}} & \multirow{2}{*}{\textbf{\#Parameter}} 
\\  & & AUC & GAUC & AUC & GAUC & AUC & GAUC & AUC & GAUC & AUC & GAUC & AUC & GAUC & AUC & GAUC & AUC & GAUC\\
\midrule
\multirow{4}{*}{\makecell{HoME w/o fg2}} & 1 & 77.83 & 72.27 & 83.14 & 77.36 & 73.89 & 69.92 & 97.00 & 85.17 & 92.58 & 78.95 & 93.09 & 80.69 & 92.72 & 77.31 & 95.61 & 84.80 & 268Mil \\
\multirow{4}{*}{} & 2 & \textbf{77.85} & \textbf{72.30} & \textbf{83.16} & \textbf{77.39} & \textbf{73.94} & \textbf{69.95} & \textbf{97.02} & \textbf{85.19} & \textbf{92.60} & \textbf{78.99} & 93.10 & \textbf{80.71} & \textbf{92.74} & \textbf{77.34} & \textbf{95.62} & \textbf{84.83} & 268Mil   \\
\multirow{4}{*}{} & 4 & 77.84 & 72.29 & \textbf{83.16} & 77.37 & 73.91 & 69.94 & 97.01 & 85.18 & 92.59 & 78.97 & \textbf{93.11} & \textbf{80.71} & 92.73 & \textbf{77.34} & 95.61 & 84.80 & 268Mil      \\
\multirow{4}{*}{} & 6 & 77.84 & 72.28 & 83.15 & \textbf{77.39} & 73.91 & 69.94 & 97.00 & 85.17 & 92.59 & 78.96 & 93.10 & \textbf{80.71} & 92.73 & 77.32 & 95.61 & 84.82 & 268Mil     \\

\bottomrule
\end{tabular}
}
\label{ablationstudylora}
\end{table*}

\begin{table*}[t]
\footnotesize
\centering
\setlength{\tabcolsep}{7pt}{
\caption{Online A/B testing results of Short-Video services at Kuaishou.}
\vspace{-1mm}
\begin{tabular}{l|c|ccccccccc}
\toprule
\multirow{2}{*}{Applications} &\multirow{2}{*}{Groups}                    & \multicolumn{3}{c}{Watching-Time Metrics}                    & \multicolumn{6}{c}{Interaction Metrics}                                                                    \\
\cmidrule(r){3-5}  \cmidrule(r){6-11}
                                     & &Average Play-time & Play-time   & Video View                  & Click                     & Like                   & Comment                  & Collect & Forward    &Follow                 \\
\midrule
\multirow{2}{*}{\makecell{Kuaishou\\Single Page}} & Young People  & \textbf{+0.770\%}        & \textbf{+1.041\%}         & \textbf{+0.547\%}                  & \textbf{-}         & \textbf{+1.036\%}         & \textbf{+2.124\%} & \textbf{+2.048\%}     & \textbf{+2.390\%}  & \textbf{+2.741\%}   \\
\multirow{2}{*}{} & Total    & \textbf{+0.311\%}         & \textbf{+0.636\%}         & \textbf{+0.059\%}                  & \textbf{-}         & \textbf{+0.601\%}         & \textbf{+1.966\%} & \textbf{+0.548\%}  & \textbf{+2.008\%} & \textbf{+1.351\%} \\
\midrule
\multirow{2}{*}{\makecell{Kuaishou Lite\\Single Page}} & Young People  & \textbf{+0.512\%}         & \textbf{+0.729\%}         & -0.215\%                  & \textbf{-}         & \textbf{+0.198\%}         & \textbf{+1.533\%} & \textbf{+1.049\%}    & \textbf{+5.241\%}   & \textbf{+1.910\%}   \\
\multirow{2}{*}{} & Total    & \textbf{+0.474\%}         & \textbf{+0.735\%}         & -0.173\%                  & \textbf{-}         & \textbf{+0.192\%}         & \textbf{+1.726\%} & \textbf{+0.856\%}  & \textbf{+2.245\%} & \textbf{+1.366\%}  \\
\midrule

\multirow{2}{*}{\makecell{Kuaishou\\Double Page}} & Young People  & \textbf{+0.311\%}         & \textbf{+0.645\%}         & \textbf{+0.498\%}     & \textbf{-}         & \textbf{+1.175\%}         & \textbf{+3.244\%} & \textbf{+1.209\%}    & \textbf{+0.717\%}  & \textbf{+0.882\%}    \\
\multirow{2}{*}{} & Total    & \textbf{+0.169\%}         & \textbf{+1.283\%}         & \textbf{+0.882\%}       & \textbf{+0.945\%}         & \textbf{+0.483\%}         & \textbf{+0.495\%} & \textbf{+1.678\%}  & \textbf{+0.795\%}  & \textbf{+0.911\%}   \\

\bottomrule
\end{tabular}
\label{mainonline}
}
\end{table*}

\subsection{Discussion of HoME Situation}
\begin{figure*}[t!]
\begin{center}
\includegraphics[width=18cm,height=7.5cm]{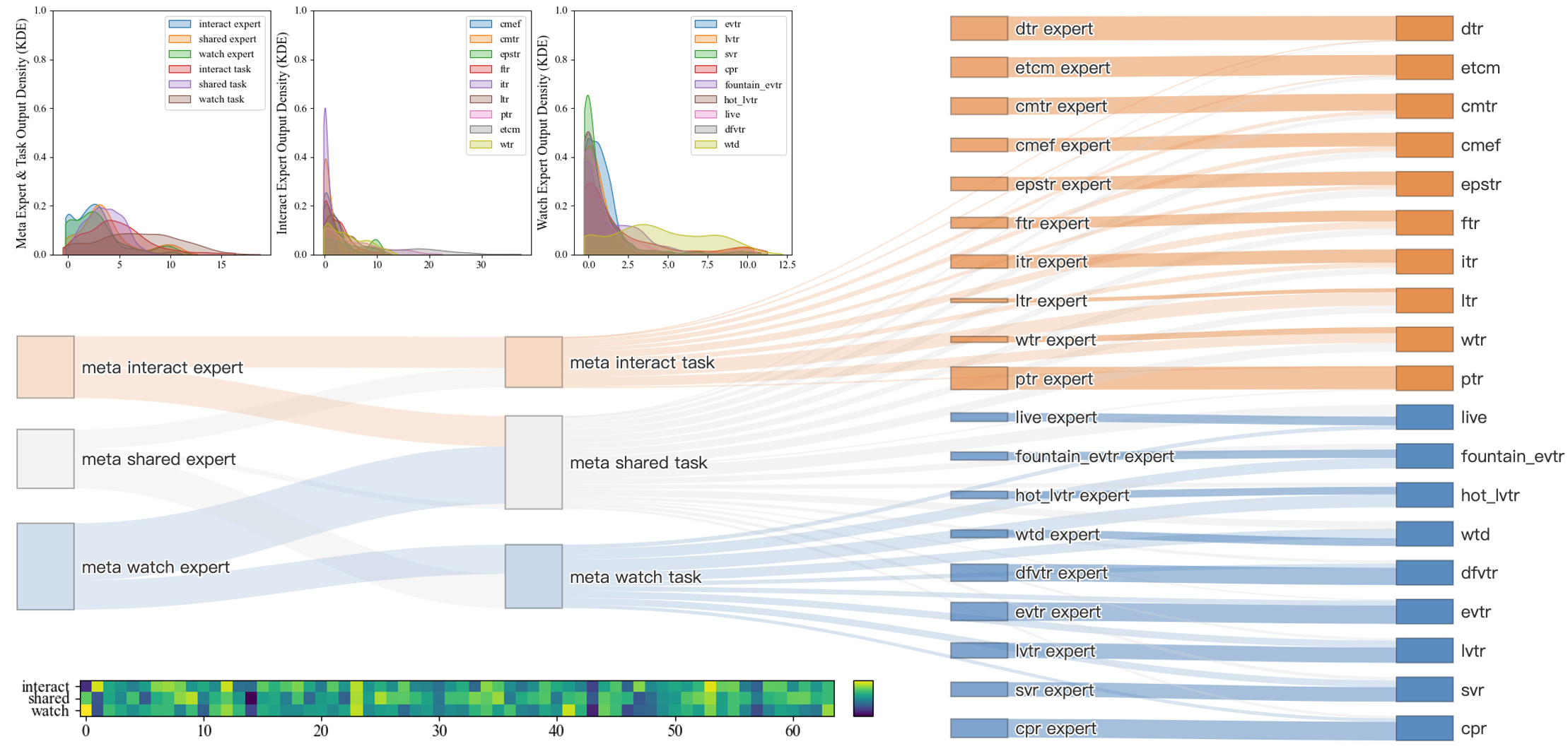}
\caption{The feature-gate heatmap, expert output distributions and gate weights flow of our HoME.}
\label{situation}
\end{center}
\end{figure*}

Figure~\ref{situation} gives the expert output distributions and graph weights flow of our HoME.
From it, we can observe that HoME achieves a balanced gate weight equilibrium situation that:
(1) According to the heatmap of feature-gate (randomly visualized 64 dimensions), we can draw a conclusion that our feature-gate could achieve a flexible element-wise feature selection for each expert.
(2) All shared and specific expert outputs are aligned in similar numerical magnitude. Further, we can find the meta-shared-expert distributions are different from specific-expert distributions, which indicates the shared-knowledge tends to be encoded by meta networks while the difference-knowledge is pushed to be encoded by the specific experts.
(3) All experts play their expected roles; the shared and specific experts contribute perceivable weights.

\subsection{Online A/B Test}
In this section, we also push HoME to be an online ranking model served at three short-video scenarios: Kuaishou Single/Double Page (in Figure~\ref{kuaishou}) and Kuaishou Lite Single Page.
In our service, the main metric is the watching-time metrics, e.g., (average) play-time, which reflects the total amount of time users spend on Kuaishou, and we also show the video view metric, which measures the total amount of short-video users watched.
The online A/B test results of Young and Total user groups are shown in Table.~\ref{mainonline}.
Actually, the about 0.1\% improvement in play-time is a statistically significant enough modification at Kuaishou. Our proposed HoME achieves a very significant improvement of $+0.311\%$, $+0.474\%$ and $+0.169\%$ for all users in three scenarios respectively, which is the most remarkable modification in the past year.
In addition, we can observe that HoME achieves significant business gain at all interaction metrics, e.g., Click, Like, Comment and others, which reveals HoME could converge the multi-task system to a more balanced equilibrium state without the seesaw phenomenon.
Moreover, we can find that the increase is larger for sparse behavior tasks, which indicates our HoME enables all shared or specific experts to obtain appropriate gradients to maximize its effectiveness.

\section{Conclusions}
In this paper, we focus on solving the multi-task learning methods practical problems and lessons we learned from Kuaishou short-video service, which is one of the world's largest recommendation scenarios.
We first figure out that the existing wide-used multi-task family, i.e., Gated Mixture-of-Expert, is prone to several serious problems that limit the model's expected ability.
From the expert outputs, we find the expert collapse problem that experts' output distributions are significantly different.
From the shared-expert learning, we observe the expert degradation problem that some shared experts only serve one task.
From the specific-expert learning, we noticed the expert underfitting problem that some sparse tasks specific-experts almost do not contribute any information.
To overcome them, we propose three insightful improvements:
(1) the Expert normalization\&Swish mechanism to align expert output distribution;
(2) the Hierarchy mask mechanism to regularize the relationship between tasks to maximize shared-expert efficiency;
(3) the Feature-gate and Self-gate mechanisms to privatize more flexible experts' inputs and connect adjacent related experts to ensure all experts could obtain appropriate gradients.
Furthermore, via extensive offline and online experiments on one of the world's largest short-video platforms, Kuaishou, we showed that HoME has led to substantial improvements compared to other wide-used multi-task methods.
Our HoME has been widely deployed on various online models at Kuaishou, supporting several services for 400 Million active users daily.

\section{Biography}
\textbf{Xu Wang} is currently a researcher at Kuaishou Technology (KStar Talent Program), Beijing, China.
He received his M.S. degree from Harbin Institute of Technology, Shenzhen, China. His main research interests include recommendation systems and multi-task learning.

\noindent \textbf{Jiangxia Cao} is currently a researcher at Kuaishou Technology (KStar Talent Program), Beijing, China.
He received his Ph.D. degree from the Institute of Information Engineering, Chinese Academy of Sciences, Beijing, China. His research focuses on industry recommender and low-resource large models.
He has published over 20 papers in top-tier international conferences and journals including SIGIR, WSDM, CIKM, ACL and so on.

\noindent\textbf{Zhiyi Fu} is currently a researcher at Kuaishou Technology (KStar Talent Program), Beijing, China.
He received his M.S. and B.S. degree from Peking University, Beijing, China. His main research interests include user long-term interest modeling and multi-task learning.

\balance
\bibliographystyle{ACM-Reference-Format}
\bibliography{sample-base-extend.bib}

\end{document}